\documentclass[aps,preprint,epsfig,rotate]{revtex4}
\usepackage{graphicx}
\usepackage{bm}
\usepackage{epsfig}

\input epsf
\begin{document}
\title{\bf Optical activity tensor for radiating atomic and molecular
       systems}

 \author{Alexei M. Frolov}
 \email[E--mail address: ]{afrolov@uwo.ca}

 \author{David M. Wardlaw}
 \email[E--mail address: ]{dwardlaw@uwo.ca}

\affiliation{Department of Chemistry\\
 University of Western Ontario, London, Ontario N6H 5B7, Canada}

\date{\today}

\begin{abstract}

The optical activity tensor (OAT) is explicitly derived. It is shown that to
evaluate a large number of effects related to optical activity of some
atomic/molecular system at arbitrary frequency $\omega$ of the incident
light, one needs to know only four optical activity tensors which have
twelve irreducible (tensor) components. An additional amplification factor
contains one $3 \times 3$ tensor of light scattering with three irreducible
components. The explicit dependence of all irreducible components of OAT
upon $\omega$ and some molecular parameters is derived and discussed. We
apply OAT to explain the dispersion of optical rotation in dilute solutions
of organic molecules. This study opens a new avenue in application of
methods of modern Quantum Electrodynamics to the optical activity.

PACS number(s): 33.55.+b and 33.20.Ni
\end{abstract}
\maketitle

\section{Introduction}

Our goal in this study is to derive the explicit expression for the optical
activity tensor \cite{Fro1} (below, OAT, for short) and for all its
irreducible components. It is shown that this tensor depends upon some
molecular parameters and the frequency $\omega$ of the incident light. In
general, the optical activity tensor must describe a large number of optical
phenomena directly related to the optical activity in various molecules,
quasi-molecules and many-atomic clusters. Furthermore, if we know the OAT
for some molecule, then we can predict the dispersion of molecular optical
rotation and the circular dichroism for this molecule. It is clear $a$
$priori$ that such a tensor describes the corrections to `regular' light
scattering \cite{Pla}. These corrections correspond to the lowest order
approximation in terms of an expansion in the fine structure constant
$\alpha$. In high order approximations one needs to apply mathematical
constructions which are significantly more complicated than regular $3
\times 3$ tensors. In this study we perform a detailed analysis of the
tensor of optical activity \cite{Fro1}. In particular, we derive the
formulas for each of the irreducible components of the tensor of optical
activity and discuss their $\omega-$dependencies.

The following approach is taken here. We consider the scattering of a photon
by a system of electrons, which is referred below as a molecule. The
molecule has two (discrete) energy levels designated below as state 1 and
state 2. The scattering of the photon means that the initial photon $({\bf
k}, \omega)$ is absorbed by the molecule. Simultaneously another photon
$({\bf k}^{\prime}, \omega^{\prime})$ is emitted by this molecule. Finally,
the molecule may either stay in the same (initial) state, or it can be
transferred into a state which is different from its initial state. In the
first case we are dealing with the non-shifted light scattering (or Rayleigh
scattering). In the second case the light frequency changes by the value
$\omega^{\prime} - \omega = E_1 - E_2$, where $E_1$ and $E_2$ are the
initial and final energies of the molecule. It is clear that in this case we
consider the shifted light scattering (also called the Raman scattering).

As is well known from quantum-electrodynamics (see, e.g., \cite{Grein},
\cite{LLQE}) the operator of electromagnetic perturbation has no matrix
elements for transitions in which two photon occupation numbers change
simultaneously. The scattering effects appear only in the second
approximation of perturbation theory. The exact formulas will be given
below, but here we want to describe the general features of our solution. In
second order perturbation theory we need to determine the matrix element
$V_{21}$ of the transition 1 $\rightarrow$ 2, i.e. from the incident
molecular state 1 into the final molecular state 2. The explicit formula for
this matrix elements is
\begin{equation}
 V_{21} = \sum{}^{\prime} \Bigl( \frac{V^{\prime}_{2n} V_{n1}}{{\cal E} -
 E_n} + \frac{V_{2n} V^{\prime}_{n1}}{{\cal E} - E_n - \omega -
 \omega^{\prime}} \Bigr)
\end{equation}
where ${\cal E} = E_1 + \omega$ is the total energy of the system which
consists of the molecule and radiation quanta $({\bf k}, \omega)$. Also, in
this formula $V_{...}$ are the matrix elements for the absorbtion of the
photon $({\bf k}, \omega)$, while $V^{\prime}_{...}$ are the matrix elements
for the emission of the photon $({\bf k}^{\prime}, \omega^{\prime})$. By
computing the matrix element $V_{21}$ we can determine the differential
cross-section
\begin{equation}
 d\sigma = \mid V_{21} \mid^2 \frac{(\omega^{\prime})^2 do^{\prime}}{4
 \pi^2} \label{eq03}
\end{equation}
where $do^{\prime}$ is a solid-angle element for the direction of the final
photon $({\bf k}^{\prime}, \omega^{\prime})$. Our first goal below is to
derive the analytical formulas for the differential cross-section $d\sigma$.
At the next step we want to investigate the dependencies of this
cross-section upon different parameters of the system and the frequency of
the incident light.

\section{The tensor of light scattering}

In this study we develop an approach which is based on the method used
earlier by G. Placzek \cite{Pla}. In \cite{Pla} Placzek introduced the
tensor of light scattering $(C_{ik})_{21}$ and derived the following formula
for the differential scattering cross-section of light by an
atomic/molecular system
\begin{eqnarray}
 d\sigma = \frac{\omega (\omega + \omega_{12})^3}{\hbar^2 c^4} \mid
 (C_{ik})_{21} ({\bf e}^{\prime}_i)^{*} {\bf e}_k \mid^2 do^{\prime}
 \label{eq1}
\end{eqnarray}
where $(C_{ik})_{21}$ is the 3 $\times$ 3 tensor of light scattering, while
${\bf e}^{\prime}_i$ and ${\bf e}_k$ are the polarization vectors of the
incident and final photons, respectively. The angular variables
$do^{\prime}$ in Eq.(\ref{eq1}) are the angular variables of the final
photon which is designated by the superscript ${}^{\prime}$, i.e.
$do^{\prime} = sin\theta^{\prime} d\theta^{\prime} d\phi$. The light
scattering tensor $(C_{ik})_{21}$ in Eq.(\ref{eq1}) is \cite{LLQE}
\begin{eqnarray}
 (C_{ik})_{21} = \sum_{n} \Bigl[ \frac{(d_i)_{2n} (d_k)_{n1}}{\omega_{n1} -
 \omega -\imath 0} + \frac{(d_k)_{2n} (d_i)_{n1}}{\omega_{n1} +
 \omega^{\prime} -\imath 0} \Bigr] \label{eq2}
\end{eqnarray}
where $\omega^{\prime} = \omega + \omega_{12}$, while $d_i$ and $d_k$ are
the corresponding components of the vector of the dipole moment ${\bf d}$.
The inifnitesimal imaginary increments in the denominators correspond to the
regular rule for pole avoidance in quantum-mechanical perturbation theory
(for more detail, see, e.g., \cite{LLQE}). Note that the differential
cross-section $d \sigma$, Eq.(\ref{eq1}), corresponds to the lowest order
approximation in terms of an expansion in the fine structure constant
$\alpha \approx \frac{1}{137}$ and contains only contributions from the
electric dipole-dipole interaction.

The tensor of light scattering $(C_{ik})_{21}$ in Eq.(\ref{eq1}) can be
written as the sum of its irreducible components $(C_{ik})_{21} = C^{0}_{21}
+ (C^{s}_{ik})_{21} + (C^{a}_{ik})_{21}$, where $C^{0}_{21}$ is the scalar
component, $(C^{s}_{ik})_{21}$ is the symmetric tensor component with zero
trace and $(C^{a}_{ik})_{21}$ is the antisymmetric tensor component. The
expressions for these irreducible components are:
\begin{eqnarray}
 C^{0}_{21} = \frac13 \sum^3_{i=1} (C_{ii})_{21} =
 \frac13 \sum_n \frac{(\omega_{n1} + \omega_{n2}) ({\bf d}_{2n} \cdot {\bf
 d}_{n1})}{(\omega_{n1} - \omega) (\omega_{n2} + \omega)}  \label{scal} \\
 (C^{s}_{ik})_{21} = \frac12 \sum_n \frac{\omega_{n1} +
 \omega_{n2}}{(\omega_{n1} - \omega) (\omega_{n2} + \omega)}
 \bigl[({\bf d}_{2n})_{i} ({\bf d}_{n1})_{k} + ({\bf d}_{2n})_{k}
 ({\bf d}_{n1})_{i} \Bigr] - C^{0}_{21} \delta_{ik} \label{symm} \\
 (C^{a}_{ik})_{21} = \frac{2 \omega + \omega_{12}}{2}
 \sum_n \frac{\bigl[({\bf d}_{2n})_{i} ({\bf d}_{n1})_{k} -
 ({\bf d}_{2n})_{k} ({\bf d}_{n1})_{i} \Bigr]}{(\omega_{n1} - \omega)
 (\omega_{n2} + \omega)}
 \label{asymm}
\end{eqnarray}
where ${\bf a}_i$ is the $i$-th (Cartesian) component of the ${\bf a}$
vector. The formulas Eqs.(\ref{scal}) - (\ref{asymm}) allow one to consider
the $\omega-$dependence (or dispersion, for short) of many properties
related to light scattering. Note that these expressions follow the work of
Placzek \cite{Pla} which was a fundamental contribution to the
Quantum-Mechanical Dispersion Theory (QMDT) developed earlier by Kramers and
Heisenberg \cite{KraHei} and Dirac \cite{Dir}. Since 1934 the approach
proposed by Placzek \cite{Pla} was widely and successfully used in
applications to various atoms and molecules. In this study we consider the
lowest order correction (in terms of the expansion in $\alpha$) to the
leading (Placzek) term defined in QMDT.

\section{The plane waves with zero spatial dispersion}

The Placzek approach for atoms suggests attempting to derive an analogous
method for molecules which would describe their optical activity. In this
Section this problem is considered in detail and it is shown that, in the
lowest order approximation, the optical activity can be described by
product of the tensor $(C_{ik})_{21}$ of light scattering and the four (or
two in some cases) new tensors. These tensors are called the tensors of
molecular optical activity, or optical activity tensors (OAT), for short.
For a regular atom in its ground state such a product of the $(C_{ik})_{21}$
tensor with the tensor of molecular optical activity equals zero
identically, and this explains the word `molecular' in the old definition of
such a tensor. Note that in some actual cases the four/two tensors of
optical activity are reduced to one tensor only.

To produce the closed analytical expressions for the optical activity
tensors in this study we shall assume that the electromagnetic field is
represented as a combination of plane waves. Each of these plane waves has
its own frequency $\omega$ and polarization which is represented by the
vector ${\bf e}$. The wave functions of the incident and final photons can
be taken in the form (see, e.g., \cite{LLQE})
\begin{eqnarray}
 {\bf A}_{{\bf e} \omega} = \sqrt{\frac{2 \pi}{\omega}} exp(-\imath \omega
 t + \imath {\bf k} \cdot {\bf r}) {\bf e}  \; \; \; , \; \; \;
 {\bf A}_{{\bf e}^{\prime} \omega^{\prime}} = \sqrt{\frac{2
 \pi}{\omega^{\prime}}} exp(-\imath \omega^{\prime} t + \imath {\bf
 k}^{\prime} \cdot {\bf r}^{\prime}) {\bf e}^{\prime} \; \; \; , \label{eq8}
\end{eqnarray}
where $\omega$ and $\omega^{\prime}$ are the corresponding frequencies,
while vectors ${\bf e}$ and ${\bf e}^{\prime}$ represent the polarization of
the incident (absorbed) and final (emitted) photons, respectively. Note that
for the emitted photon we need its conjugate wave function, i.e.
${\bf A}^{*}_{{\bf e}^{\prime} \omega^{\prime}}$, rather than the wave
function defined in Eq.(\ref{eq8}). Below, we shall consider the plane waves
in the transverse (or radiation) gauge, where $div {\bf A} = 0$. In this
gauge one finds ${\bf k} \cdot {\bf e} = 0$ and ${\bf k}^{\prime} \cdot {\bf
e}^{\prime} = 0$. As follows from these equations the electric ${\bf E}$ and
magnetic ${\bf H}$ fields are
\begin{eqnarray}
 {\bf E}_{{\bf e} \omega} = - \frac{\partial}{\partial t} {\bf A}_{{\bf e}
 \omega} = -\imath \sqrt{2 \pi \omega} {\bf e} exp(-\imath \omega t +
 \imath {\bf k} \cdot {\bf r}) \label{eq40} \\
 {\bf H}_{{\bf e} \omega} = curl {\bf A}_{{\bf e} \omega} = \imath
 \sqrt{\frac{2 \pi}{\omega}} ({\bf k} \times {\bf e}) exp(-\imath \omega t
 + \imath {\bf k} \cdot {\bf r}) \label{eq41}
\end{eqnarray}
By introducing the unit vector ${\bf n} = \frac{{\bf k}}{\omega}$ we can
re-write the last equation in the form
\begin{equation}
 {\bf H}_{{\bf e} \omega} = \imath \sqrt{2 \pi \omega} ({\bf n} \times
 {\bf e}) exp(-\imath \omega t + \imath {\bf k} \cdot {\bf r})
\end{equation}
Analogous expressions can be obtained for the ${\bf E}_{{\bf e}^{\prime}
\omega^{\prime}}$ and ${\bf H}_{{\bf e}^{\prime} \omega^{\prime}}$ fields
\begin{eqnarray}
 {\bf E}_{{\bf e}^{\prime} \omega^{\prime}} = \frac{\partial}{\partial t}
 {\bf A}^{*}_{{\bf e}^{\prime} \omega^{\prime}} = \imath \sqrt{2 \pi
 \omega^{\prime}} {\bf e} exp(\imath \omega^{\prime} t - \imath
 {\bf k}^{\prime} \cdot {\bf r}^{\prime}) \\
 {\bf H}_{{\bf e}^{\prime} \omega^{\prime}} = curl {\bf A}^{*}_{{\bf
 e}^{\prime} \omega^{\prime}} = -\imath \sqrt{2 \pi \omega^{\prime}}
 ({\bf n} \times {\bf e}^{*}) exp(\imath \omega^{\prime} t - \imath
 {\bf k}^{\prime} \cdot {\bf r}^{\prime})
\end{eqnarray}

From these equations one finds the following expressions for the electric
dipole and magnetic dipole interactions. In fact, for each of the (${\bf e},
\omega$)-components of the ${\bf E}$ and ${\bf H}$ vectors we have
\begin{equation}
 V^{e}_{{\bf e} \omega} = -{\bf d} \cdot {\bf E}_{{\bf e} \omega} =
 \imath \sqrt{2 \pi \omega} ({\bf d} \cdot {\bf e}) exp(-\imath \omega t
 + \imath {\bf k} \cdot {\bf r})
\end{equation}
and
\begin{equation}
 V^{m}_{{\bf e} \omega} = -{\bf m} \cdot {\bf H}_{{\bf e} \omega} =
 - \imath \sqrt{2 \pi \omega} [{\bf m} \cdot ({\bf n} \times {\bf e})]
 exp(-\imath \omega t + \imath {\bf k} \cdot {\bf r})
\end{equation}
where ${\bf d}$ and ${\bf m}$ are the vectors of the electric and magnetic
dipole moments, respectively. For the emitted photon we have analogously
\begin{equation}
 (V^{e})^{\prime}_{{\bf e}^{\prime} \omega^{\prime}} = -{\bf d} \cdot
 {\bf E}_{{\bf e}^{\prime} \omega^{\prime}} = -\imath \sqrt{2 \pi
 \omega^{\prime}} ({\bf d} \cdot ({\bf e}^{\prime})^{*}) exp(\imath
 \omega^{\prime} t - \imath {\bf k}^{\prime} \cdot {\bf r}^{\prime})
 \label{dipl}
\end{equation}
and
\begin{equation}
 (V^{m})^{\prime}_{{\bf e}^{\prime} \omega^{\prime}} = -{\bf m} \cdot {\bf
 H}_{{\bf e}^{\prime} \omega^{\prime}} =
 \imath \sqrt{2 \pi \omega^{\prime}} \{{\bf m} \cdot [{\bf n} \times
 ({\bf e}^{\prime})^{*}]\} exp(\imath \omega^{\prime} t - \imath {\bf
 k}^{\prime} \cdot {\bf r}^{\prime})
 \label{magn}
\end{equation}
In derivation of these formulas we have assumed that the vector of dipole
moment ${\bf d}$ is real. Note that such an assumption corresponds to the
case of classical mechanics. In quantum mechanics the vector ${\bf d}$ is
always real (in the coordinate representation), while the vector ${\bf m}$
is a complex vector, since it contains an imaginary unit $\imath$ as a
factor. Furthermore, in applications of perturbation theory in quantum
electrodynamics one finds a number of additional advantages if the following
identities are obeyed $(V^{e})^{\prime} = (V^{e})^{*}$ and $(V^{m})^{\prime}
= (V^{m})^{*}$ for the interactions which involve the emitted photon(s). In
our case, this means that Eq.(\ref{dipl}) does not change its form, while in
Eq.(\ref{magn}) one finds the ${\bf m}^{*}$ vector instead of ${\bf m}$
vector
\begin{equation}
 (V^{m})^{\prime}_{{\bf e}^{\prime} \omega^{\prime}} =
 \imath \sqrt{2 \pi \omega^{\prime}} \{{\bf m}^{*} \cdot [{\bf n} \times
 ({\bf e}^{\prime})^{*}]\} exp(\imath \omega^{\prime} t - \imath {\bf
 k}^{\prime} \cdot {\bf r}^{\prime})
 \label{magn2}
\end{equation}

Now, we can introduce an approximation that the wavelengths $\lambda$ of
the incident and final photons are significantly larger than a typical
linear size $a$ of the molecule (our light scatterer). In this case we have
${\bf k} \cdot {\bf r} \le \mid {\bf k} \mid \mid {\bf r} \mid \ll
\frac{a}{\lambda} \approx 0$. In this approximation one finds from
Eqs.(\ref{eq40}) and (\ref{eq41})
\begin{equation}
 {\bf E}_{{\bf e} \omega} = -\imath \sqrt{2 \pi \omega} {\bf e}
 exp(-\imath \omega t) \; \; \; and \; \; \;
 {\bf H}_{{\bf e} \omega} = \imath \sqrt{2 \pi \omega} ({\bf n} \times
 {\bf e}) exp(-\imath \omega t) \label{magn3}
\end{equation}
Analogous expressions for the $({\bf e}^{\prime}, \omega^{\prime})$
components are
\begin{equation}
 {\bf E}_{{\bf e}^{\prime} \omega^{\prime}} = \imath \sqrt{2 \pi
 \omega^{\prime}} ({\bf e}^{\prime})^{*} exp(\imath \omega^{\prime} t)
 \; \; \; and \; \; \;
 {\bf H}_{{\bf e}^{\prime} \omega^{\prime}} = -\imath \sqrt{2 \pi
 \omega^{\prime}} [{\bf n} \times ({\bf e}^{\prime})^{*}] exp(\imath
 \omega^{\prime} t) \label{magn4}
\end{equation}

Plane waves with zero spatial dispersion are often used for analysis of
interactions between light (with relatively large wavelengths $\lambda$) and
matter. In particular, below, we shall use only these plane waves with zero
spatial dispersion. In general, this approximation is very useful for
studying of light scattering ($\lambda \geq 2000$ $\AA$) on atoms and
relatively small molecules with $a_{max} \le$ 500 - 800 $\AA$. For our
present purposes the use of this approximation is crucial, since only this
approximation allows one to derive the closed analytical expressions for all
matrix elements required in our procedure (see below).

\section{Perturbation theory}

As we mentioned above the scattering effects in an electromagnetic field
appear only in the second order perturbation theory. In this second order
approximation the matrix element $V_{21}$ for the transition between states
1 and 2 is written in the following form \cite{LLQE}
\begin{equation}
 V_{21} = \sum_{n} \Bigl( \frac{V^{\prime}_{2n} V_{n1}}{{\cal E}_1 -
 {\cal E}^{I}_n} + \frac{V_{2n} V^{\prime}_{n1}}{{\cal E}_1 -
 {\cal E}^{II}_n} \Bigr) \label{eq42}
\end{equation}
where the notation ${\cal E}$ designates the total energy of the system
(`molecule + photons'), i.e. in the case considered here we have ${\cal
E}^{I}_n = E_n$ and ${\cal E}^{II}_n = E_n + \omega + \omega^{\prime}$,
where $E_n$ is the energy of an intermediate atomic state, $\omega$ is the
frequency of the incident light and $\omega^{\prime} = \omega + \omega_{21}
= E_2 - E_1 + \omega$ is the frequency of the final photon. The
matrix element $V_{ab}$ represents absorbtion of the photon with the wave
vector ${\bf k}$. Analogously, the matrix element $V^{\prime}_{ab}$
represents emission of the photon with the wave vector ${\bf k}^{\prime}$.
In the general case, in Eq.(\ref{eq42}) the $V_{ab}$ and $V^{\prime}_{ab}$
interactions are represented in the forms $V = V^{e} + V^{m} + V^{qe} +
V^{qm} + \ldots$ and $V^{\prime} = (V^{e})^{\prime} + (V^{m})^{\prime} +
(V^{qe})^{\prime} + (V^{qm})^{\prime} + \ldots$, respectively. Here $V^{e},
V^{m}, V^{qe}$ are the electric dipole, magnetic dipole and electric
quadruple interactions, respectively. Keeping only lowest order terms in the
fine structure constant expansion of $V$, we can write for these
interactions $V \approx V^{e} + V^{m}$ and $V^{\prime} \approx
(V^{e})^{\prime} + (V^{m})^{\prime}$. In this case one finds from
Eq.(\ref{eq42})
\begin{eqnarray}
 V_{21} = \sum_{n} \Bigl[ \frac{(V^{e})^{\prime}_{2n} V^{e}_{n1}}{{\cal
 E}_1 - {\cal E}^{I}_n} +
 \frac{V^{e}_{2n} (V^{e})^{\prime}_{n1}}{{\cal E}_1 - {\cal E}^{II}_n}
 \Bigr]
 + \sum_{n} \Bigl[ \frac{(V^{e})^{\prime}_{2n} V^{m}_{n1}}{{\cal
 E}_1 - {\cal E}^{I}_n} +
 \frac{V^{e}_{2n} (V^{m})^{\prime}_{n1}}{{\cal E}_1 - {\cal E}^{II}_n}
 + \frac{(V^{m})^{\prime}_{2n} V^{e}_{n1}}{{\cal
 E}_1 - {\cal E}^{I}_n} \label{equa43} \\
 + \frac{V^{m}_{2n} (V^{e})^{\prime}_{n1}}{{\cal E}_1 - {\cal E}^{II}_n}
 \Bigr] + \sum_{n} \Bigl[ \frac{(V^{m})^{\prime}_{2n} V^{m}_{n1}}{{\cal
 E}_1 - {\cal E}^{I}_n} +
 \frac{V^{m}_{2n} (V^{m})^{\prime}_{n1}}{{\cal E}_1 - {\cal E}^{II}_n}
 \Bigr] + \ldots \nonumber
\end{eqnarray}
By neglecting here by all terms $\sim V^{m} V^{m}$ and other terms of higher
order in the fine structure constant $\alpha$, we obtain the following
formula for the differential cross-section of light scattering $d \sigma$
\begin{eqnarray}
 d\sigma = \mid V_{21} \mid^2 \frac{(\omega^{\prime})^2 do^{\prime}}{4
 \pi^2} \approx \Bigl\{ \sum_{n} \Bigl[ \frac{(V^{e})^{\prime}_{2n}
 V^{e}_{n1}}{{\cal E}_1 - {\cal E}^{I}_n} + \frac{V^{e}_{2n}
 (V^{e})^{\prime}_{n1}}{{\cal E}_1 - {\cal E}^{II}_n}\Bigr] \Bigr\}^2
 \frac{(\omega^{\prime})^2 do^{\prime}}{4 \pi^2} \nonumber \\
 + \Bigl\{ \sum_{n} \Bigl[ \frac{(V^{e})^{\prime}_{2n} V^{e}_{n1}}{{\cal
 E}_1 - {\cal E}^{I}_n} + \frac{V^{e}_{2n} (V^{e})^{\prime}_{n1}}{{\cal E}_1
 - {\cal E}^{II}_n}\Bigr] \Bigr\} \cdot \Bigl\{ \sum_{n} \Bigl[
 \frac{(V^{e})^{\prime}_{2n} V^{m}_{n1}}{{\cal E}_1 -
 {\cal E}^{I}_n} + \frac{V^{e}_{2n} (V^{m})^{\prime}_{n1}}{{\cal E}_1 -
 {\cal E}^{II}_n} \nonumber \\
 + \frac{(V^{m})^{\prime}_{2n} V^{e}_{n1}}{{\cal E}_1 -
 {\cal E}^{I}_n} + \frac{V^{m}_{2n} (V^{e})^{\prime}_{n1}}{{\cal E}_1 -
 {\cal E}^{II}_n} \Bigr] \Bigr\} \frac{(\omega^{\prime})^2 do^{\prime}}{4
 \pi^2} = d\sigma_{ee} + d\sigma_{em} \label{equ44}
\end{eqnarray}
where $d\sigma_{ee}$ is the part of the total cross-section which is reduced
to the expression given above (see Eq.(\ref{eq1})). This part of the
cross-section is not related to the optical activity. The second term in the
right-hand side of Eq.(\ref{equ44}) is significantly smaller, in the general
case, than the first term, i.e. $d\sigma_{em} \ll d\sigma_{ee}$. However,
the second term in Eq.(\ref{equ44}) is a great interest, since it represents
new physical effects, including effects directly related to the optical
activity.

As follows from Eq.(\ref{equ44}) in order to determine the part of the
total cross-section responsible for molecular optical activity in the lowest
order approximation we need to obtain the explicit formulas for the matrix
elements of $V^{e} (V^{e})^{\prime}, V^{e} (V^{m})^{\prime}, V^{m}
(V^{e})^{\prime}$ and other similar products. For the plane waves with
non-zero spatial dispersion the arising expressions are extremely
complicated, since each of the $V^{e}$ and/or $V^{m}$ interactions contains
an infinite number of $V^{e}_{{\bf e} \omega}$ and $V^{m}_{{\bf e} \omega}$
components. In general, in $V^{e} (V^{m})^{\prime}, V^{m} (V^{e})^{\prime}$
and other similar products one finds an infinite number of cross-terms which
explicitly depend upon coordinates. These terms cannot be computed without a
complete and accurate knowledge of the molecular electron density
$\rho_e({\bf r})$. In turn, this problem is almost unsolvable in the general
case. However, in the approximation $\lambda \gg a$ mentioned above we do
not need to know the molecular electron density $\rho_e({\bf r})$ in each
spatial point. Briefly, this means that we can use only the two `molecular'
vectors ${\bf d}$ and ${\bf m}$, which are assumed to be constant in any
spatial point inside of the molecule. These two vectors ${\bf d}$ and ${\bf
m}$ are the vectors of electric dipole and magnetic dipole moment,
respectively. The vector of the electric dipole moment ${\bf d}$ is a polar
vector, while the magnetic dipole moment ${\bf m}$ is a axial vector (or
pseudo-vector).

The approximation $\lambda \gg a$ corresponds to the use of the plane waves
with zero spatial dispersion defined in the end of previous Section. By
using these plane waves we can write for $V^{e}, (V^{e})^{\prime}, V^{m}$
and $(V^{m})^{\prime}$
\begin{equation}
 V^{e}_{{\bf e} \omega} = \imath \sqrt{2 \pi \omega} ({\bf d} \cdot {\bf
 e}) exp(-\imath \omega t) \; \; \; , \; \; \;
 (V^{e})^{\prime}_{{\bf e}^{\prime} \omega^{\prime}} = -\imath \sqrt{2 \pi
 \omega} ({\bf d} \cdot {\bf e}^{*}) exp(\imath \omega t)
\end{equation}
and
\begin{equation}
 V^{m}_{{\bf e} \omega} = -\imath \sqrt{2 \pi \omega} [{\bf m} \cdot
 ({\bf n} \times {\bf e})] exp(-\imath \omega t) \; \; \; , \; \; \;
 (V^{m})^{\prime}_{{\bf e}^{\prime} \omega^{\prime}} = \imath \sqrt{2 \pi
 \omega^{\prime}} \{{\bf m}^{*} \cdot [{\bf n} \times ({\bf
 e}^{\prime})^{*}]\} exp(\imath \omega t)
\end{equation}
where ${\bf d}$ and ${\bf m}$ are the corresponding vectors of the dipole
and magnetic moments of the molecule.

The next step of our procedure is to derive the explicit formula for the
cross-section from which the formula for the optical activity tensor (OAT)
will follow directly. This is our goal in the next Section.

\section{The tensor of molecular optical activity}

By using the expressions derived in the previous Sections we can write the
following formula for the differential cross-section $d\sigma_{em}$ (defined
above in Eq.(\ref{equ44}))
\begin{eqnarray}
 d\sigma_{em} = \Big| \sum_{n} \frac{({\bf d}_{2n} \cdot {\bf e}^{\prime})
 ({\bf d}_{n1} \cdot {\bf e})}{\omega_{n1} - \omega - \imath 0} +
 \frac{({\bf d}_{2n} \cdot {\bf e}) ({\bf d}_{n1} \cdot {\bf
 e}^{\prime})}{\omega_{n1} + \omega^{\prime} - \imath 0} \Bigr|
 \Big| \sum_{n} \frac{({\bf d}_{2n} \cdot {\bf e}^{\prime})
 [({\bf m}_{n1} \times {\bf n}) \cdot {\bf e}]}{\omega_{n1} - \omega -
 \imath 0} + \label{eq21} \\
 \frac{[({\bf m}^{*}_{2n} \times {\bf n}) \cdot {\bf e}^{\prime}] ({\bf
 d}_{n1} \cdot {\bf e})}{\omega_{n1} - \omega - \imath 0} +
 \frac{({\bf d}_{2n} \cdot {\bf e})
 [({\bf m}^{*}_{n1} \times {\bf n}) \cdot {\bf e}^{\prime}]}{\omega_{n1} +
 \omega^{\prime} - \imath 0}
 + \frac{[({\bf m}_{2n} \times {\bf n}) \cdot {\bf e}] ({\bf d}_{n1} \cdot
 {\bf e}^{\prime})}{\omega_{n1} + \omega^{\prime} - \imath 0} \Bigr| \cdot
 \frac{\omega (\omega^{\prime})^3}{\hbar^2 c^4} do^{\prime} \nonumber
\end{eqnarray}
where the notation ${\bf e}^{\prime}$ designates the vector $({\bf
e}^{\prime})^{*}$. This system of notation is also used in the two following
equations. The equation Eq.(\ref{eq21}) can be re-written as
\begin{eqnarray}
 d\sigma_{em} = \Big| \sum_{n} \frac{({\bf d}_{2n} \cdot {\bf e}^{\prime})
 ({\bf d}_{n1} \cdot {\bf e})}{\omega_{n1} - \omega - \imath 0} +
 \frac{({\bf d}_{2n} \cdot {\bf e}) ({\bf d}_{n1} \cdot {\bf
 e}^{\prime})}{\omega_{n1} + \omega^{\prime} - \imath 0} \Bigr|
 \Big| \sum_{n} \frac{({\bf d}_{2n} \cdot {\bf e}^{\prime})
 [{\bf m}_{n1} \cdot ({\bf n} \times {\bf e})]}{\omega_{n1} - \omega -
 \imath 0} + \label{cross1} \\
 \frac{[{\bf m}^{*}_{2n} \cdot ({\bf n} \times {\bf e}^{\prime})] ({\bf
 d}_{n1} \cdot {\bf e})}{\omega_{n1} - \omega - \imath 0} +
 \frac{({\bf d}_{2n} \cdot {\bf e})
 [{\bf m}^{*}_{n1} \cdot ({\bf n} \times {\bf e}^{\prime})]}{\omega_{n1} +
 \omega^{\prime} - \imath 0}
 + \frac{[{\bf m}_{2n} \cdot ({\bf n} \times {\bf e})] ({\bf d}_{n1} \cdot
 {\bf e}^{\prime})}{\omega_{n1} + \omega^{\prime} - \imath 0} \Bigr| \cdot
 \frac{\omega (\omega^{\prime})^3}{\hbar^2 c^4} do^{\prime} \nonumber
\end{eqnarray}
Note that the vector ${\bf n}$ in these equations corresponds to the
direction of the scattered light. Formally, this vector can be oriented in
an arbitrary spatial direction, but in almost all modern experiments on
optical activity in homogeneous solutions the direction of the scattered
light always coincides with the direction of the incident light. This means
that our differential cross-section must be multiplied by a delta-function
$\delta({\bf n}_{in} - {\bf n})$ and integrated over the angular variables
$o^{\prime} = (\theta^{\prime}, \phi^{\prime})$ of the unit vector ${\bf n}
= (cos\theta^{\prime} cos\phi^{\prime}, cos\theta^{\prime} sin\phi^{\prime},
sin\theta^{\prime})$ which represents the direction of the final photon. The
unit vector ${\bf n}_{in}$ describes the direction of the incident photon.
This produces the following expression for the cross-section $\sigma_{em}$
\begin{eqnarray}
 \sigma_{em} = \frac{4 \pi \omega (\omega + \omega_{12})^3}{\hbar^2 c^4}
 \cdot \Big| \sum_{n} \frac{({\bf d}_{2n} \cdot {\bf e}^{\prime})
 ({\bf d}_{n1} \cdot {\bf e})}{\omega_{n1} - \omega - \imath 0} +
 \frac{({\bf d}_{2n} \cdot {\bf e}) ({\bf d}_{n1} \cdot {\bf
 e}^{\prime})}{\omega_{n1} + \omega^{\prime} - \imath 0} \Bigr|
 \times \nonumber \\
 \Big| \sum_{n} \frac{({\bf d}_{2n} \cdot {\bf e}^{\prime})
 [{\bf m}_{n1} \cdot ({\bf n}_{in} \times {\bf e})]}{\omega_{n1} - \omega
 - \imath 0}
 + \frac{[{\bf m}^{*}_{2n} \cdot ({\bf n}_{in} \times {\bf e}^{\prime})]
 ({\bf d}_{n1} \cdot {\bf e})}{\omega_{n1} - \omega - \imath 0} +
 \frac{({\bf d}_{2n} \cdot {\bf e})
 [{\bf m}^{*}_{n1} \cdot ({\bf n}_{in} \times {\bf
 e}^{\prime})]}{\omega_{n1} + \omega^{\prime} - \imath 0} \nonumber \\
 + \frac{[{\bf m}_{2n} \cdot ({\bf n}_{in} \times {\bf e})] ({\bf d}_{n1}
 \cdot {\bf e}^{\prime})}{\omega_{n1} + \omega^{\prime} - \imath 0} \Bigr|
 \label{cross3}
\end{eqnarray}
where $\omega^{\prime} = \omega + \omega_{12}$ and unit-vector ${\bf
n}_{in}$ designates the direction of propagation of the incident photon.

The expression, Eq.(\ref{cross3}), can be cast in the following form
\begin{eqnarray}
 \sigma_{em} = \frac{4 \pi \omega (\omega + \omega_{12})^3}{\hbar^2 c^4}
 \cdot \Bigl| (C_{ik})_{21} ({\bf e}^{\prime}_i)^{*} {\bf e}_k \Bigr| \cdot
 \Bigl| (S_{ik})_{21} ({\bf e}^{\prime})^{*}_i ({\bf n}_{in} \times
 {\bf e})_k + (T_{ik})_{21} ({\bf n}_{in} \times ({\bf e}^{\prime})^{*})_i
 ({\bf e})_k \nonumber \\
 + (U_{ik})_{21} {\bf e}_i ({\bf n}_{in} \times ({\bf
 e}^{\prime})^{*})_k + (V_{ik})_{21} ({\bf n}_{in} \times {\bf
 e})_i ({\bf e}^{\prime})^{*}_k \Bigr| \label{main}
\end{eqnarray}
where $(S_{ik})_{21}, (T_{ik})_{21}, (U_{ik})_{21}$ and $(V_{ik})_{21}$ are
$3 \times 3$ electro-magnetic dipole-dipole tensors, while the electric
dipole-dipole tensor $(C_{ik})_{21}$ is defined above in Eq.(\ref{eq1}). The
explicit formulas for these tensors are
\begin{eqnarray}
 (S_{ik})_{21} = \frac{({\bf d}_{2n})_{i} ({\bf m}_{n1})_{k}}{\omega_{n1}
 - \omega} = \frac{(d_i)_{2n} (m_k)_{n1}}{\omega_{n1}
 - \omega} \\
 (U_{ik})_{21} = \frac{({\bf m}^{*}_{2n})_{i} ({\bf
 d}_{n1})_{k}}{\omega_{n1} - \omega} = \frac{(m_i)^{*}_{2n}
 (d_k)_{n1}}{\omega_{n1} - \omega} \\
 (T_{ik})_{21} = \frac{({\bf d}_{2n})_{i} ({\bf
 m}^{*}_{n1})_{k}}{\omega_{n1} + \omega^{\prime}} = \frac{(d_i)_{2n}
 (m_k)^{*}_{n1}}{\omega_{n1} + \omega^{\prime}} \\
 (V_{ik})_{21} = \frac{({\bf m}_{2n})_{i} ({\bf d}_{n1})_{k}}{\omega_{n1}
 + \omega^{\prime}} = \frac{(m_i)_{2n} (d_k)_{n1}}{\omega_{n1}
 + \omega^{\prime}}
\end{eqnarray}
Here we assume that, in the general case, the vectors ${\bf e}^{\prime}$ and
${\bf e}$ which represent the polarization of light are complex. Note that
in the last equations and everywhere below we drop the infinitesimal
imaginary increments in the denominators which indicate the avoidance of the
poles in the $\omega-$plane. Each of these tensors can be represented as a
sum of its irreducible components, e.g., $S_{ik} = S^{0} \delta_{ik} +
S^{s}_{ik} + S^{a}_{ik}$, where
\begin{equation}
 S^{0} = \frac13 \sum^3_{i=1} S_{ii} \; \; \; , \; \; \;
 S^{s}_{ik} = \frac12 (S_{ik} + S_{ki}) - S^{0} \delta_{ik} \; \; \; , \; \;
 \; S^{a}_{ik} = \frac12 (S_{ik} - S_{ki})
\end{equation}
are the scalar, symmetric tensor and antisymmetric tensor components of the
tensor $S_{ik}$, respectively. The scalar, symmetric and antisymmetric
tensor components of the $T_{ik}, U_{ik}, V_{ik}$ tensors are defined
analogously. Note that all components of these irreducible $S^{0}, T^{0},
U^{0}, V^{0}, S^{s}_{ik}, T^{s}_{ik}, U^{s}_{ik}, V^{s}_{ik}, S^{a}_{ik},
T^{a}_{ik}, U^{a}_{ik}$ and $V^{a}_{ik}$ tensors contain the products of
different components of the ${\bf d}$ and ${\bf m}$ vectors, which are the
vectors of the electric dipole moment and magnetic dipole moment,
respectively. The vector-operator which represents the electric dipole
moment is assumed to be self-conjugate. In general, this is true only in
the coordinate representation. For instance, the explicit expressions for
the $S^{0}, T^{0}, U^{0}$ and $V^{0}$ scalars (they are also called the
scalar-components of the $S, T, U$ and $V$ tensors) are
\begin{eqnarray}
 (S^{0})_{21} = \frac13 \sum_{n} \frac{{\bf d}_{2n} \cdot {\bf
 m}_{n1}}{\omega_{n1} - \omega} \; \; \; , \; \; \; \;
 (T^{0})_{21} = \frac13 \sum_{n} \frac{{\bf m}^{*}_{2n} \cdot
 {\bf d}_{n1}}{\omega_{n1} - \omega} \; \; \; , \label{tens} \\
 (U^{0})_{21} = \frac13 \sum_{n} \frac{{\bf d}_{2n} \cdot
 {\bf m}^{*}_{n1}}{\omega_{n2} + \omega} \; \; \; , \; \; \; \;
 (V^{0})_{21} = \frac13 \sum_{n} \frac{{\bf m}_{2n} \cdot
 {\bf d}_{n1}}{\omega_{n2} + \omega} \; \; \; , \nonumber
\end{eqnarray}
respectively. The formulas for the symmetric and antisymmetric parts of the
$S, T, U$ and $V$ tensors are slightly more complicated. For instance, for
the $(S^{s}_{ik})_{21}$ and $(S^{a}_{ik})_{21}$ tensors one finds the
following formulas
\begin{eqnarray}
 (S^{s}_{ik})_{21} = \frac12 \sum_{n} \frac{(d_i)_{2n} (m_k)_{n1} +
 (d_k)_{2n} (m_i)_{n1}}{\omega_{n1} - \omega} - (S^{0})_{21}
 \delta_{ik} \label{sym2} \\
 (S^{a}_{ik})_{21} = \frac12 \sum_{n} \frac{(d_i)_{2n} (m_k)_{n1} -
 (d_k)_{2n} (m_i)_{n1}}{\omega_{n1} - \omega} \label{asym2}
\end{eqnarray}
The symmetric and antisymmetric tensor components of the $T, U$ and $V$
tensors have can be written in a very similar form
\begin{eqnarray}
 (T^{s}_{ik})_{21} = \frac12 \sum_{n} \frac{(m_i)^{*}_{2n} (d_k)_{n1} +
 (m_k)^{*}_{2n} (d_i)_{n1}}{\omega_{n1} - \omega} - (T^{0})_{21}
 \delta_{ik} \label{sym3} \\
 (T^{a}_{ik})_{21} = \frac12 \sum_{n} \frac{(m_i)^{*}_{2n} (d_k)_{n1} -
 (m_k)^{*}_{2n} (d_i)_{n1}}{\omega_{n1} - \omega} \label{asym3} \\
 (U^{s}_{ik})_{21} = \frac12 \sum_{n} \frac{(d_i)_{2n} (m_k)^{*}_{n1} +
 (d_k)_{2n} (m_i)^{*}_{n1}}{\omega_{n2} + \omega} - (U^{0})_{21}
 \delta_{ik} \label{sym4} \\
 (U^{a}_{ik})_{21} = \frac12 \sum_{n} \frac{(d_i)_{2n} (m_k)^{*}_{n1} -
 (d_k)_{2n} (m_i)^{*}_{n1}}{\omega_{n2} + \omega} \label{asym4} \\
 (V^{s}_{ik})_{21} = \frac12 \sum_{n} \frac{(m_i)_{2n} (d_k)_{n1} +
 (m_k)_{2n} (d_i)_{n1}}{\omega_{n2} + \omega} - (V^{0})_{21}
 \delta_{ik} \label{sym5} \\
 (V^{a}_{ik})_{21} = \frac12 \sum_{n} \frac{(m_i)_{2n} (d_k)_{n1} -
 (m_k)_{2n} (d_i)_{n1}}{\omega_{n2} + \omega} \label{asym5}
\end{eqnarray}
These expressions explicitly define each of the components of the optical
activity tensor (OAT). The formulas derived in this Section allow one to
describe a large number of optical phenomena related to the optical activity
in various molecular, atomic and quasi-atomic systems. Some applications of
these formulas in the case of Rayleigh light scattering are discussed in the
next Section.

\section{Rayleigh light scattering. Dispersion of the optical rotation}

In the case of Rayleigh light scattering the formulas given above are
simplified substantially, since the incident state is identical with the
final state, i.e. in these formulas we need to use $1 = 2, \omega_{12} = 0,
\omega^{\prime} = \omega$, etc. The tensor of light scattering,
Eq.(\ref{eq2}), now has only two irreducible components: (1) the scalar
component
\begin{equation}
 C^{0}_{11} = \frac23 \sum_n \frac{\omega_{n1}}{(\omega^2_{n1} -
 \omega^2)} ({\bf d}_{1n} \cdot {\bf d}_{n1}) \label{scalr}
\end{equation}
and symmetric (tensor) component
\begin{equation}
 (C^{s}_{ik})_{11} = \sum_n \frac{\omega_{n1}}{(\omega^2_{n1} - \omega^2)}
 \bigl[({\bf d}_{1n})_{i} ({\bf d}_{n1})_{k} + ({\bf d}_{1n})_{k}
 ({\bf d}_{n1})_{i} \Bigr] - C^{0}_{11} \delta_{ik} \label{symmr}
\end{equation}
The third (or antisymmetric) irreducible component of the tensor of light
scattering $(C^{a}_{ik})_{11}$ equals zero identically, since the
vector-operator of the dipole moment ${\bf d}$ is self-conjugate and each
of its components is real, i.e. we can write $(d_i)_{1n} (d_k)_{n1} =
(d_k)_{1n} (d_i)_{n1}$ and, therefore, $(C^{a}_{ik})_{11} = 0$.

The irreducible components of the tensor of molecular optical activity take
the following form
\begin{eqnarray}
 (S^{0})_{11} = \frac13 \sum_{n} \frac{{\bf d}_{1n} \cdot {\bf
 m}_{n1}}{\omega_{n1} - \omega} \; \; \; , \; \; \; \;
 (T^{0})_{11} = \frac13 \sum_{n} \frac{{\bf m}^{*}_{1n} \cdot
 {\bf d}_{n1}}{\omega_{n1} - \omega} \; \; \; , \label{tens33} \\
 (U^{0})_{11} = \frac13 \sum_{n} \frac{{\bf d}_{1n} \cdot
 {\bf m}^{*}_{n1}}{\omega_{n1} + \omega} \; \; \; , \; \; \; \;
 (V^{0})_{11} = \frac13 \sum_{n} \frac{{\bf m}_{1n} \cdot
 {\bf d}_{n1}}{\omega_{n1} + \omega} \nonumber
\end{eqnarray}
for the scalar components, and
\begin{eqnarray}
 (S^{s}_{ik})_{11} = \frac12 \sum_{n} \frac{(d_i)_{1n} (m_k)_{n1} +
 (d_k)_{1n} (m_i)_{n1}}{\omega_{n1} - \omega} - (S^{0})_{11}
 \delta_{ik} \label{sym33} \\
 (S^{a}_{ik})_{11} = \frac12 \sum_{n} \frac{(d_i)_{1n} (m_k)_{n1} -
 (d_k)_{1n} (m_i)_{n1}}{\omega_{n1} - \omega} \label{asym33} \\
 (T^{s}_{ik})_{11} = \frac12 \sum_{n} \frac{(m_i)^{*}_{1n} (d_k)_{n1} +
 (m_k)^{*}_{1n} (d_i)_{n1}}{\omega_{n1} - \omega} - (T^{0})_{11}
 \delta_{ik} \label{sym333} \\
 (T^{a}_{ik})_{11} = \frac12 \sum_{n} \frac{(m_i)^{*}_{1n} (d_k)_{n1} -
 (m_k)^{*}_{1n} (d_i)_{n1}}{\omega_{n1} - \omega} \label{asym333} \\
 (U^{s}_{ik})_{11} = \frac12 \sum_{n} \frac{(d_i)_{1n} (m_k)^{*}_{n1} +
 (d_k)_{1n} (m_i)^{*}_{n1}}{\omega_{n1} + \omega} - (U^{0})_{11}
 \delta_{ik} \label{sym433} \\
 (U^{a}_{ik})_{11} = \frac12 \sum_{n} \frac{(d_i)_{1n} (m_k)^{*}_{n1} -
 (d_k)_{1n} (m_i)^{*}_{n1}}{\omega_{n1} + \omega} \label{asym433} \\
 (V^{s}_{ik})_{11} = \frac12 \sum_{n} \frac{(m_i)_{1n} (d_k)_{n1} +
 (m_k)_{1n} (d_i)_{n1}}{\omega_{n1} + \omega} - (V^{0})_{11}
 \delta_{ik} \label{sym533} \\
 (V^{a}_{ik})_{11} = \frac12 \sum_{n} \frac{(m_i)_{1n} (d_k)_{n1} -
 (m_k)_{1n} (d_i)_{n1}}{\omega_{n1} + \omega} \label{asym533}
\end{eqnarray}
for the symmetric and anti-symmetric tensor components, respectively. These
formulas determine each component of the optical activity tensor in the case
of Rayleigh light scattering, i.e. for non-shifted light scattering
$\omega_{12} = 0$.

Let us apply these formulas to the dilute solution of organic substances. In
modern organic chemistry the researcher routinely measures the so-called
optical rotation, i.e. the angle by which the plane of linearly polarized
light is turned about the direction of light propagation as the light
travels through dilute solutions of organic substances. Theory of optical
rotation by chiral organic molecules is a well developed area of theoretical
chemistry (see, e.g., \cite{Mason}). The fundamental formulas and results in
this theory were derived to the middle of 1940's \cite{EWC}, \cite{Rose}
(more references and discussion can be found in \cite{Mason} and
\cite{Baron}). In all these works the approach based on the direct solution
of Maxwell equations have been used. The optical activity tensor for
radiating atomic and/or molecular systems was not constructed. If our new
approach is correct, then we must obtain the same formulas and results. We
consider the optical rotation by chiral organic molecules as an example for
application of our theory.

In general, the optical rotation is measured by comparing orientation of the
plane of linearly polarized light before and after its propagation through a
solution which contains chiral organic molecules. In such experiments the
directions of propagation of the incident and final light coincide with each
other. Therefore, we can apply the formulas derived above. Formally, the
rotation of the plane of linearly polarized light is described by the factor
(or scalar product) ${\bf n} \cdot ({\bf e}_i \times {\bf e}^{\prime}_i) =
{\bf e}^{\prime}_i \cdot ({\bf n} \times {\bf e}_i)$, where the unit vector
${\bf n}$ determines the direction of light propagation, while the unit
vectors ${\bf e}_i$ and ${\bf e}^{\prime}_i$ are the vectors which describe
polarization of the incident and final light. In the case of linearly
polarized light all these vectors are real. The angle between these two
vectors (${\bf e}_i$ and ${\bf e}^{\prime}_i$) determines the `optical
rotation' of the plane of linearly polarized light. If there are no chiral
molecules in the solution, then we have ${\bf n} \cdot ({\bf e}_i \times
{\bf e}^{\prime}_i) = {\bf n} \cdot ({\bf e}_i \times {\bf e}_i) = 0$, i.e.
no optical rotation at all.

The expression for the cross-section, Eq.(\ref{main}) in the case of pure
optical rotation is modified to the form
\begin{eqnarray}
 \sigma_{em} = \frac{4 \pi \omega^4}{\hbar^2 c^4}
 \cdot \Bigl| (C^0)_{21} ({\bf e}^{\prime} \cdot {\bf e}) \Bigr| \cdot
 \Bigl| (S^0)_{21} {\bf n}_{in} \cdot ({\bf e} \times {\bf e}^{\prime}) -
 (T^0)_{21} {\bf n}_{in} \cdot ({\bf e} \times {\bf e}^{\prime}) \nonumber \\
 - (U^0)_{21} {\bf n}_{in} \cdot ({\bf e} \times {\bf e}^{\prime}) +
   (V^0)_{21} {\bf n}_{in} \cdot ({\bf e} \times {\bf e}^{\prime}) \Bigr|
 \label{main1}
\end{eqnarray}
This formula can be re-arranged to the form
\begin{eqnarray}
 \sigma_{em} = \frac{4 \pi \omega^4}{\hbar^2 c^4}
 \cdot \Bigl| (C^0)_{21} \Bigr| \cdot \Bigl| (S^0)_{21} - (T^0)_{21}
 - (U^0)_{21} + (V^0)_{21} \Bigr| \cdot \Bigl| ({\bf e}^{\prime} \cdot
 {\bf e}) \Bigr| \cdot \Bigl| {\bf n}_{in} \cdot ({\bf e} \times
 {\bf e}^{\prime}) \Bigr| \label{main2}
\end{eqnarray}
By using the known expressions for all scalar components of the tensors
(see, Eqs.(\ref{scalr}) and (\ref{tens33}) above) one finds
\begin{eqnarray}
 \sigma_{em} = \frac{4 \pi \omega^4}{\hbar^2 c^4}
 \Bigl| \frac23 \sum_n \frac{\omega_{n1}}{(\omega^2_{n1} -
 \omega^2)} ({\bf d}_{1n} \cdot {\bf d}_{n1}) \Bigr|
 \Bigl| \frac23 \sum_n \frac{1}{\omega_{n1} -
 \omega} Im ({\bf m}_{1n} \cdot {\bf d}_{n1}) \nonumber \\
 + \frac23 \sum_n \frac{1}{\omega_{n1} + \omega} Im ({\bf d}_{1n} \cdot
 {\bf m}_{n1}) \Bigr| \Bigl| ({\bf e}^{\prime} \cdot
 {\bf e}) \Bigr| \cdot \Bigl| {\bf n}_{in} \cdot ({\bf e} \times
 {\bf e}^{\prime}) \Bigr| \label{main3}
\end{eqnarray}
where the notation $Im$ designates the imaginary part of the terms written
in the following brackets.

Assuming that always ${\bf d}_{1n} \cdot {\bf m}_{n1} = {\bf m}_{1n} \cdot
{\bf d}_{n1}$, we obtain the final formula for the $\sigma_{em}$
cross-section
\begin{eqnarray}
 \sigma_{em} = \frac{32 \pi \omega^4}{9 \hbar^2 c^4}
 \Bigl| \sum_n \frac{\omega_{n1}}{(\omega^2_{n1} - \omega^2)} ({\bf d}_{1n}
 \cdot {\bf d}_{n1}) \Bigr|
 \Bigl| \sum_n \frac{\omega_{n1}}{(\omega^2_{n1} - \omega^2)}
 Im ({\bf d}_{1n} \cdot {\bf m}_{n1}) \Bigr|
 \Bigl| ({\bf e}^{\prime} \cdot
 {\bf e}) \Bigr| \cdot \Bigl| {\bf n}_{in} \cdot ({\bf e} \times
 {\bf e}^{\prime}) \Bigr| \label{main4}
\end{eqnarray}
Note that the factor ${\bf n}_{in} \cdot ({\bf e} \times {\bf e}^{\prime})$
is the optical rotation itself. Furthermore, the scalar product $({\bf
e}^{\prime} \cdot {\bf e})$ is a constant factor if the two unit vectors
${\bf e}$ and ${\bf e}^{\prime}$ are known (they describe polarization of
the incident and final light waves). Therefore, instead of the $\sigma_{em}$
cross-section we can consider the coefficient of optical rotation (or
optical rotation, for short) $R$
\begin{eqnarray}
 R = \frac{32 \pi \omega^4}{9 \hbar^2 c^4}
 \Bigl| \sum_n \frac{\omega_{n1}}{(\omega^2_{n1} - \omega^2)} ({\bf d}_{1n}
 \cdot {\bf d}_{n1}) \Bigr| \Bigl| \sum_n \frac{\omega_{n1}}{(\omega^2_{n1}
 - \omega^2)} Im ({\bf d}_{1n} \cdot {\bf m}_{n1}) \Bigr| \label{RRR}
\end{eqnarray}
This formula follows from rigorous QED analysis and it describes the
dispersion of optical rotation $R(\omega)$ in various solutions of different
organic molecules. As follows from the formula, Eq.(\ref{RRR}), that the
optical rotation $R(\omega) \rightarrow 0$, if the frequency $\omega
\rightarrow 0$. Note that the formula Eq.(\ref{RRR}) formally works at
arbitrary light frequencies, including frequencies which correspond to the
optical transitions, i.e. to the absorbtion and/or emission of light quanta.
However, in such cases one needs to operate with the complete formulas,
which include small imaginary increments in the denominators, for the light
scattering tensor and optical activity tensor(s). For frequencies which
correspond to the optical transitions (or resonance frequencies) and for
frequencies close to them the explicit expressions for these small imaginary
increments become important. This essentially makes impossible to conclude
our universal analysis, since such increments depend upon the natural widths
of spectral lines $\Gamma_{1n} (= \Gamma_{n1})$. For different molecular
systems such natural widths are very different and they also depend on some
other factors.

In actual experiments in organic chemistry chemists measure the optical
rotatory parameter $\beta$ (angle) which is also called the chiral response
parameter (see, e.g., \cite{Djer}, \cite{Mason})
\begin{eqnarray}
 \beta = \frac{c}{6 \pi \hbar} \sum_n \frac{Im\Bigl({\bf d}_{1n} \cdot {\bf
 m}_{n1} \Bigr)}{\nu^2_{n1} - \nu^2} \label{Rosen}
\end{eqnarray}
where the summation is taken over all intermediate states. In this equation
we use the linear frequencies $\nu$ instead of circular frequencies
$\omega$, where $\omega = 2 \pi \nu$. Note that in some works the definition
of $\beta$ contains an additional factor $\sim \nu^2$ (see, e.g.,
\cite{EWC}). The formula, Eq.(\ref{Rosen}), is based on the semi-classical
analysis of optical rotation performed by Rosenfeld in \cite{Rose}. In
actual cases this formula can be applied only in very limited intervals of
light frequencies $\nu$ or wavelengths $\lambda$. In general, it cannot be
used in the `resonance areas', i.e. at the frequencies where the
absorbtion/emission of light is very high. Furthermore, to describe the
results of real experiments, even at regular frequencies, the formula
Eq.(\ref{Rosen}) is often modified with the use of some additional
`empirical' parameters.

The piece which corresponds to Eq.(\ref{Rosen}) can easily be recognized in
our Eq.(\ref{RRR}). Note that the formula Eq.(\ref{Rosen}) represents the
optical rotation, but some additional (and important) factors are missing
from Eq.(\ref{Rosen}). In particular, Eq.(\ref{Rosen}) does not contain any
component of the tensor of light scattering. Therefore, based only on this
equation we cannot conclude that frequencies, for which the absorbtion of
light is high, are also responsible very large optical rotations of
polarized light. This fact is well known from numerous experiments and it
follows from Eqs.(\ref{cross3}) and (\ref{main}). This means that
Eq.(\ref{Rosen}) cannot correctly describe the dispersion of optical
rotation at arbitrary light frequencies. Finally, we want to emphasize that
optical rotation is only one of a large number of phenomena related to the
optical activity in various atomic and molecular systems. In this Section we
have applied our method to the solutions of chiral organic molecules. The
goal of this application was to show that it produces formulas and results
which are similar (but more complete and general) to the expressions found
in earlier studies \cite{EWC}, \cite{Rose} (see also \cite{Mason},
\cite{Baron} and references therein) based on the direct solution of the
Maxwell equations.

However, we want to emphasize that there are many other atomic and molecular
systems which have a great interest for solution of many important physical
problems. Recently, a large number of such problems was discovered in
Stellar Astrophysics (see, e.g, \cite{Polar}, \cite{Jon1} and references
therein). Currently, there are many various Stars and other Stellar Objects
that emit light with noticeable partial linear and/or circular polarization,
e.g., the hot Be-stars and Herbig Ae-stars \cite{Jon1}, \cite{Petr}, the
Wolf-Rayet Stars \cite{Polar}, \cite{Mof}, etc. For many of these problems
even preliminary expressions and results were never derived due to their
extremely high complexity. The formulas derived in this study allow one to
describe the emission of polarized light by some Stellar objects. It will be
considered in our future studies.

\section{The polarization parameters for the elementary light scattering}

Let us assume that the incident light has a linear polarization which is
described by the unit polarization vector ${\bf e}$, Eq.(\ref{magn3}). After
the scattering the light also has a linear polarization which is represented
by the unit vector ${\bf e}^{\prime}$, Eq.(\ref{magn4}). In order to
describe the change of the polarization of light during its scattering we
can define the three following parameters
\begin{eqnarray}
 s_0 = \mid {\bf e} \cdot {\bf e}^{\prime} \mid^2 +
 \mid {\bf e}_{\perp} \cdot {\bf e}^{\prime} \mid^2 =
 \mid {\bf e} \cdot {\bf e}^{\prime} \mid^2 +
 \mid ({\bf e} \times {\bf n}) \cdot {\bf e}^{\prime} \mid^2 \\
 s_1 = \mid {\bf e} \cdot {\bf e}^{\prime} \mid^2 -
 \mid {\bf e}_{\perp} \cdot {\bf e}^{\prime} \mid^2 =
 \mid {\bf e} \cdot {\bf e}^{\prime} \mid^2 -
 \mid ({\bf e} \times {\bf n}) \cdot {\bf e}^{\prime} \mid^2 \\
 s_2 = 2 ({\bf e} \cdot {\bf e}^{\prime})) [({\bf e}_{\perp} \cdot
 {\bf e}^{\prime})] = 2 ({\bf e} \cdot {\bf e}^{\prime})
 [({\bf n} \times {\bf e}) \cdot {\bf e}^{\prime})]
\end{eqnarray}
where ${\bf e}_{\perp} = ({\bf n} \times {\bf e})$ is the unit vector which
represents another polarization vector which is orthogonal to the vector
${\bf e}$ and ${\bf n}$ is the direction of the incident light propagation.
Let us introduce the angle $\theta$ such that
\begin{equation}
 ({\bf e} \cdot {\bf e}^{\prime}) = cos\theta
\end{equation}
In this case, one finds
\begin{equation}
 ({\bf e}_{\perp} \cdot {\bf e}^{\prime}) = sin\theta =
 ({\bf n} \times {\bf e}) \cdot {\bf e}^{\prime} =
 {\bf n} \cdot ({\bf e} \times {\bf e}^{\prime})
\end{equation}

With these expressions we can write for the $s_0, s_1$ and $s_2$ parameters
$s_0 = 1, s_1 = cos(2 \theta)$ and $s_2 = sin(2 \theta)$. It is clear that
$s^2_0 + s^2_1 + s^2_2 = 2$. These three parameters describe polarization of
an arbitrary linearly polarized light which is represented as a combination
of plane waves with zero spatial dispersion. In some sense, the $s_0, s_1$
and $s_2$ parameters can be considered as the Stokes parameters (see, e.g.,
\cite{Polar}) defined for plane waves with zero spatial dispersion. Note
also that the absolute value of the $s_2$ parameter is included in our
formula for the $\sigma_{em}$ cross-section, Eq.(\ref{main4}).

\section{Conclusion}

We have shown that all phenomena related to the optical activity can
completely be described with the use of only four tensors: $S_{21}, T_{21},
U_{21}$ and $V_{21}$. The fifth tensor $C_{21}$ (the tensor of
electric-dipole light scattering) is included in the formula for the
cross-section $\sigma_{em}$ as an amplification factor. These five tensors
have fifteen irreducible tensor-components $C^{0}, C^{s}_{ik}, C^{a}_{ik},
S^{0}, T^{0}, U^{0}, V^{0}, S^{s}_{ik}, T^{s}_{ik}, U^{s}_{ik}, V^{s}_{ik},
S^{a}_{ik}, T^{a}_{ik}, U^{a}_{ik}$ and $V^{a}_{ik}$. The first three
tensors $C^{0}, C^{s}_{ik}, C^{a}_{ik}$ here have nothing to do with the
optical activity itself. Instead they determine the amplification factor
which also depends upon $\omega$. The optical activity is described by the
twelve tensors ($S^{0}, T^{0}, U^{0}, V^{0}, S^{s}_{ik}, T^{s}_{ik},
U^{s}_{ik}, V^{s}_{ik}, S^{a}_{ik}, T^{a}_{ik}, U^{a}_{ik}$ and
$V^{a}_{ik}$). In many real applications, however, the total number of
independent tensors can be reduced. For instance, if the $1$- and $2$-states
are identical and $\omega_{21} = 0$ (Rayleigh scattering), then to describe
optical activity one needs only two tensors (not four!) with six irreducible
components. Furthermore, if the polarization vectors are chosen as real (not
complex), then to describe the optical activity one needs only one $3 \times
3$ tensor with three irreducible components. However, the explicit
$\omega-$dependence of such a tensor will be quite complicated.

In conclusion, we wish to note that the intensity of the scattered light
$I^{\prime}$ is uniformly related to the intensity of the incident light $I$
by the relation
\begin{equation}
  I^{\prime} = \Bigl(\frac{\omega^{\prime}}{\omega}\Bigr) \sigma I
  \label{last}
\end{equation}
where $\omega^{\prime} = \omega + \omega_{12}$. As follows from the formula
for the cross-section $\sigma$, Eq.(\ref{cross3}), in any optically active
solution the direction of maximal light intensity of the (scattered) light
will always be rotated during its propagation along the direction ${\bf
n}_{in}$. The factor $\Bigl(\frac{\omega^{\prime}}{\omega}\Bigr) \sigma$ in
the last formula can be considered as the rotation power. As follows from
Eq.(\ref{last}) and Eq.(\ref{main}) the uniform combination of the twelve
irreducible tensors mentioned above multiplied by the amplification factor
$\sim \Bigl| (C_{ik})_{21} \Bigr|$ allows one to determine the so-called
rotation power of any given optically active solution. It is also clear that
the approach described above produces the complete (and correct) formula for
the $\omega-$dependence of the rotation power. Thus, we have shown that our
method produces the correct results for dilute solutions of chiral organic
molecules. Our next goal is to apply this method to various problems from
Stellar Astrophysics which are related with the emission of polarized light
\cite{FroJon}. In general, these problems are significantly more complicated
than analysis of the optical rotation in dilute solutions of chiral organic
molecules. Very likely, that our approach developed in this study will be
modified and improved before its applications to the problems from Stellar
Astrophysics. In particular, it can be necessary to replace the plane waves
with zero spatial dispersion by some more complicated functions.

\begin{center}
   {\bf Acknowledgments}
\end{center}

It is a pleasure to acknowledge the University of Western Ontario for
financial support.

\end{document}